\documentclass[10pt,showpacs,amsmath,amssymb]{article}
\usepackage{amsfonts}
\usepackage{mathdots}
\usepackage{amsmath}
\usepackage{amsthm}
\usepackage{color}

\newcommand{\field}[1]{\mathbb{#1}}

\numberwithin{equation}{section}

\begin{document}

\title{Holomorphic solutions of the susy grassmannian $\sigma$-model and gauge invariance}

\author{\vspace{1cm}\\
         {\bf V. Hussin}$^{1,2}$
        \thanks{E-mail address:
        hussin@dms.umontreal.ca} \,,
        {\bf M. Lafrance}$^{2}$
        \thanks{E-mail address:
        marie.lafrance@umontreal.ca} \,,
         {\bf \.{I}. Yurdu\c{s}en}$^{3,1}$
        \thanks{E-mail address:
        yurdusen@hacettepe.edu.tr}
         {\,, and \bf W. J. Zakrzewski}$^4$
        \thanks{E-mail address:
       w.j.zakrzewski@durham.ac.uk}
\\
\\$^1$Centre de Recherches Math\'{e}matiques, 
Universit\'{e} de Montr\'{e}al,\\ CP 6128, Succ. Centre-Ville, Montr\'{e}al, 
Quebec H3C 3J7, Canada
\\
\\$^2$D\'{e}partement de Math\'{e}matiques et de Statistique, 
Universit\'{e} de Montr\'{e}al,\\ CP 6128, Succ. Centre-Ville, Montr\'{e}al, 
Quebec H3C 3J7, Canada
\\
   \\$^3$Department of Mathematics, Hacettepe University,
                     \\ 06800 Beytepe, Ankara, Turkey
\\                     
   \\$^4$Department of Mathematical Sciences, University of Durham, 
                       \\Durham DH1 3LE, United Kingdom}




\date{\today}

\maketitle

\begin{abstract}
For the first time we develop the gauge invariance of the supersymmetric grassmannian sigma 
model $G(M,N)$. It is richer then its purely bosonic submodel and we show how to use it in order to reduce some constant curvature holomorphic solutions of the model into simpler expressions.
\end{abstract}



\section{Introduction \label{intro}}
Although gauge invariance of the supersymmetric grassmannian sigma
model ({\bf susy} $\sigma$-model) $G(M,N)$ is well known \cite{witten, susyG(MN), fujii1, fujii2}, to 
our knowledge, up to now no explicit form of it had been used in an effective way 
to analyse the solutions of the model. However, as the gauge invariance of the susy $G(M,N)$ $\sigma$-model
is richer than of its purely bosonic submodel, we can exploit this invariance to construct the 
solutions of the model in a simpler form. The aim of this manuscript is to demonstrate this fact 
explicitly. In order to make it self consistent we start by reminding the reader some properties of the susy $G(M,N)$ 
$\sigma$-model \cite{susyG(MN)}: 

The susy $G(M,N)$ $\sigma$-model is defined on a two-dimensional complex superspace 
$(x_{\pm}; \theta_{\pm})$, where $(x_+, x_-)$ are local coordinates on $\field{C}$ and 
$(\theta_+, \theta_-)$ are complex odd Grassmann variables. The bosonic superfield has the 
following expansion
\begin{eqnarray}
\Phi(x_{\pm},\theta_{\pm}) =  \Phi_0(x_{\pm}) + 
i \theta_{+} \Phi_{1}(x_{\pm}) +
i \theta_{-} \Phi_{2}(x_{\pm}) - 
\theta_{+}\theta_{-} \Phi_3(x_{\pm}),
\label{intsuperbosefield}
\end{eqnarray}
where $\Phi_0$ and $\Phi_3$ are $N \times M$ bosonic complex matrices and 
$\Phi_1$ and $\Phi_2$ are $N \times M$ fermionic complex matrices. As in a 
purely bosonic model \cite{Zbook}, the superfield $\Phi$ satisfies 
\begin{eqnarray}
\Phi^{\dagger} \Phi = I_M.  \label{norm}
\end{eqnarray}
The energy action functional of the model is given by
\begin{eqnarray}
 S = 2 \int_{S^2} dx_+d x_- d{\theta_+}d{\theta_-} \textrm{Tr} \left( |\check{D}_+\Phi|^2 - |\check{D}_-\Phi|^2 \right)\,,
 \label{action}
 \end{eqnarray}
where the supercovariant derivatives $\check{D}_{\pm}$ are defined by 
\begin{eqnarray}
 \check{D}_{\pm} = \check{\partial}_{\pm} - (\Phi^{\dagger} \check{\partial}_{\pm} \Phi)\,,
 \label{supcovariat}
 \end{eqnarray}
with usual superderivatives 
\begin{eqnarray}
\check{\partial}_{\pm} = -i \partial_{\theta_{\pm}} + \theta_{\pm} \partial_{{\pm}}\,, \qquad  \partial_{{\pm}} \equiv  \partial_{x_{\pm}}\,.
\label{superderivative}
\end{eqnarray}

Note that the susy operators satisfy $\check{\partial}_{\pm}^2 = -i {\partial}_{\pm} $. Using the principle of least 
action, it is found that the superfield $\Phi$ satisfies the Euler-Lagrange equations of the model
\begin{eqnarray}
\check{D}_{+} \check{D}_{-} \Phi + \Phi | \check{D}_{-} \Phi |^2 = 0\,,
\label{ELeqs}
\end{eqnarray}
together with the constraint (\ref{norm}). We know that if we want to obtain finite action solutions 
of the susy $G(M,N)$ $\sigma$-model, we have to impose additionally the boundary conditions $\check{D}_{\pm} \Phi \rightarrow 0$, 
$|x_{\pm}| \rightarrow \infty$. 

In the following section we use the well known MacFarlane parametrization (first used for finding solutions of the purely
bosonic models \cite{MacFarlane}, and later used also in the susy case \cite{MacFarlanesusy}) and look at the solutions of the 
susy $G(M,N)$ $\sigma$-model. In Section \ref{gaugeinv}, we show 
that these susy solutions can be effectively transformed into a simpler form by explicitly using 
the full gauge invariance of the model. In Section \ref{examples}, two examples are presented 
to emphasize the effectiveness of using the gauge transformation in the context of 
susy $G(M,N)$ $\sigma$-model.  Finally, we give some concluding remarks and discussions of our ongoing 
projects.

\section{MacFarlane parametrization and solutions of the susy $G(M,N)$ $\sigma$-model  \label{macfarlane}}
Let us start with a holomorphic solution of the susy $G(M,N)$ $\sigma$-model, written as
\begin{eqnarray}
W(x_+, \theta_+) = Z(x_+) + i \theta_+ \eta(x_+) A(x_+)\,,
\label{MFparasusy}
\end{eqnarray}
where we have introduced a fermionic function $\eta(x_+)$ so that the matrices 
$Z \in \field{C}^{N \times M}$ and $A \in \field{C}^{N \times M}$ are both usual bosonic 
matrix functions of $x_+$. Moreover, the form of $Z$ is given by 
\begin{eqnarray}
Z = \left( \begin{array}{c}
I_M \\
K
\end{array} \right) \,, \qquad K \in \field{C}^{(N - M) \times M} \,.
\label{zMF}
\end{eqnarray}

The form (\ref{zMF}) guarantees that $Z(x_+)$ is written in the usual MacFarlane parametrization ({\it i.e.}; if we drop the fermionic part of (\ref{MFparasusy}) we end with the holomorphic solution of the purely bosonic model). At this stage 
$A(x_+)$ is arbitrary and a natural question that can be asked is: Would it be possible to 
simplify the form of (\ref{MFparasusy}) by exploiting the gauge invariance of the susy model? 

Let us first observe that the solution $W $ given by  (\ref{MFparasusy}) satisfies 
$\check{\partial}_- W = 0$, where $\check{\partial}_- $ is defined in (\ref{superderivative}). 

This form of the solution is equivalent to a solution of the original susy model (\ref{ELeqs}) given by 
\begin{eqnarray}
\Phi(x_{\pm}, \theta_{\pm}) = W(x_+, \theta_+) L(x_{\pm}, \theta_{\pm})\,,
\label{gensolsusy}
\end{eqnarray}
where $L(x_{\pm}, \theta_{\pm})$ is an $M \times M$ matrix. The function $\Phi$ has to satisfy the condition
\begin{eqnarray}
\Phi^{\dagger} \Phi = I_M \quad \iff \quad W^{\dagger} W = (L L^{\dagger})^{-1}\,.
\label{constraintofmodel}
\end{eqnarray}

Proving that the expression for $\Phi(x_{\pm}, \theta_{\pm})$ given in (\ref{gensolsusy}),
which comes from the holomorphic expression $W(x_+, \theta_+)$ solves (\ref{ELeqs}) is easy; here one repeats the steps used
in the proof for the purely bosonic case. 
In order to see that 
$\check{\partial}_- W = 0$ implies $\check{D}_{-} \Phi = 0$, we
consider 
\begin{eqnarray}
\check{D}_{-} \Phi &=& \check{\partial}_{-} \Phi - \Phi (\Phi^{\dagger} \check{\partial}_{-} \Phi)\,, \nonumber \\
&=& (1 - \Phi \Phi^{\dagger}) \check{\partial}_{-} \Phi\,, \nonumber \\ 
&=& W (\check{\partial}_{-} L) - \Phi \Phi^{\dagger} W (\check{\partial}_{-} L)\,, \nonumber \\
&=& W (\check{\partial}_{-} L) - W (L L^{\dagger}) (W^{\dagger} W) (\check{\partial}_{-} L)\,\,=\,0,
\label{proofholomorph}
\end{eqnarray}
due to (\ref{constraintofmodel}).

In the purely bosonic case, this is the whole story; in the susy case $L$ is an $M\times M$ matrix superfield and so it makes
sense to exploit this freedom to get a simpler expression for the additional matrix $A$. 
So let us look at the relations between $L$ and $A$ and, in particular, the implications of the right hand side of (\ref{constraintofmodel}). Looking first at (\ref{MFparasusy}) we get
\begin{eqnarray}
W^{\dagger} W = Z^{\dagger} Z + i \theta_+ \eta (Z^{\dagger} A) + 
i \theta_- \eta^{\dagger} (A^{\dagger} Z) - \theta_+\theta_-\eta^{\dagger}\eta (A^{\dagger} A)
\,.
\label{proWdagW}
\end{eqnarray}    
To express the product $L L^{\dagger}$ we use the usual decomposition of a matrix 
superfield $L(x_{\pm}, \theta_{\pm})$ 
\begin{eqnarray}
L = L_0+ i \theta_+ \eta L_1 + 
i \theta_- \eta^{\dagger} L_2- \theta_+\theta_-\eta^{\dagger}\eta L_3
\,,
\label{superL}
\end{eqnarray}
together with the properties of Grassmann variables:
\begin{eqnarray}
\theta_+^2 = \theta_-^2 = 0\,, \quad \theta_+\theta_- = - \theta_-\theta_+\,, \quad 
\theta_+ \eta = - \eta \theta_+\,, \quad (\theta_+ \eta)^{\dagger} = -\eta^{\dagger}\theta_-
\,.
\label{grass}
\end{eqnarray}
This way we find the conditions which have to be imposed on the matrices $L_0$, $L_1$, $L_2$, and $L_3$ 
so that they satisfy (\ref{constraintofmodel}):
\begin{eqnarray}
&L_0 L_0^{\dagger} = M_0^{-1}\,, \label{4condL1} \\
&L_0 L_2^{\dagger} + L_1 L_0^{\dagger} = -M_0^{-1} M_1 M_0^{-1}\,, \label{4condL2} \\
&L_0 L_1^{\dagger} + L_2 L_0^{\dagger} = -M_0^{-1} M_2 M_0^{-1}\,, \label{4condL3} \\
&L_0 L_3^{\dagger} + L_3 L_0^{\dagger} + L_1 L_1^{\dagger} + L_2 L_2^{\dagger} = - M_0^{-1} (M_3 - M_1 M_0^{-1} M_2 
- M_2 M_0^{-1} M_1) M_0^{-1}\,,
\label{4condL4}
\end{eqnarray}
where
\begin{eqnarray}
M_0 = Z^{\dagger} Z\,, \quad M_1 = Z^{\dagger} A\,, \quad M_2 = A^{\dagger} Z\,, \quad M_3 = A^{\dagger} A\,. 
\label{defineM}
\end{eqnarray}

Clearly $L_0$ is obtained from the purely bosonic solution $Z$. The two equations, 
involving $L_1$ and $L_2$ only are equivalent, since $M_2^{\dagger} = M_1$. As we are looking at the conditions that
have to be imposed on the matrices $L_i$ we observe that, without any loss of generality, 
we can assume $L_1 = 0$ and $L_3 = L_3^{\dagger}$ and we will satisfy all conditions on $L$. Indeed we find that 
\begin{eqnarray}
L_2 = -M_0^{-1} M_1^{\dagger} L_0\,,
\label{L2}
\end{eqnarray}
and  
\begin{eqnarray}
L_0 L_3 + L_3 L_0 = -M_0^{-1} (M_3 - M_1 M_0^{-1} M_1^{\dagger}) M_0^{-1}\,.
\label{L3}
\end{eqnarray}

In the subsequent development we will not need the explicit form of $L$. However, the choice of 
$L_1 = 0$ will play an important role in simplifying the final expressions.

\section{Gauge invariance of the susy $G(M,N)$ $\sigma$-model 
\label{gaugeinv}}
The gauge invariance of the susy $G(M,N)$ $\sigma$-model is obtained by generalising its purely bosonic submodel 
to simplify the form of the solutions of the model. 

If $\Phi$ is a solution of (\ref{ELeqs}), then it can be shown that 
\begin{eqnarray}
\tilde{\Phi} = V \Phi U\,,
\label{gaugephi}
\end{eqnarray}
is also a solution where $V = V_0 \in U(N)$ is a constant matrix (purely bosonic) and 
$U = U(x_{\pm}, \theta_{\pm}) \in U(M)$. It is easy to see why $V_0$ has to be a constant matrix. Indeed we note 
that the supercovariant derivatives given in (\ref{supcovariat}) and the superfield 
$\Phi$ transform in a similar fashion under the gauge transformation. Hence, starting with a general 
transformation matrix $V = V (x_{\pm}, \theta_{\pm})$ the imposition of the condition that $\Phi$ and 
$\check{D}_{\pm} \Phi$ transform in the same way, we immediately obtain that $V$ is a constant matrix. 
This result is similar to the purely bosonic case \cite{Zbook}. So let us consider the conditions on $U$.

Since the matrix $U$ depends on $x_{\pm}$ and 
also on $\theta_{\pm}$, we can perform the usual decomposition 
\begin{eqnarray}
U = U_0+ i \theta_+ \eta U_1 + 
i \theta_- \eta^{\dagger} U_2- \theta_+\theta_-\eta^{\dagger}\eta U_3
\,,
\label{decomU}
\end{eqnarray}
where now $U_i = U_i (x_+, x_-) \in \field{C}^{M \times M}$, $i=0, 1, 2, 3$ and the fermionic function $\eta(x_+)$ is the same as before.

The condition of unitarity of 
$U$ may be written explicitly as
\begin{eqnarray}
U^{\dagger} U = U U^{\dagger} = I_M \iff
\left\{\begin{array}{c}
U_0^{\dagger} = U_0^{-1}\,, \quad U_1 = - U_0 U_2 ^{\dagger} U_0\,, \\
U_3 + U_0 U_1^{\dagger} U_1 + U_1 U_1^{\dagger} U_0 + U_0  U_3^{\dagger} U_0 = 0\,.
\end{array} \right.
\label{cononU}
\end{eqnarray}

Now, we address the question of how to exploit gauge invariance to reduce the solution $\Phi$ and 
thus $W$ to simpler expressions. Since we want to preserve the form of the 
purely bosonic solution $Z$, we consider the simplified solution in the form 
\begin{eqnarray}
\Phi_R = W_R L_R = (Z + i \theta_+ \eta A_R) L_R\,,
\label{reducedsoln}
\end{eqnarray}
and want to find $V = V_0$ and $U$ such that 
\begin{eqnarray}
\Phi_R = V_0 \Phi U\,, \quad \iff  W_R L_R = V_0 W L U\,.
\label{reducedsoln}
\end{eqnarray}

Writing $W_R$ and $W$ explicitly (\ref{reducedsoln}) becomes 
\begin{eqnarray}
(Z + i \theta_+ \eta A_R) L_R = V_0 (Z + i \theta_+ \eta A) L U\,.
\label{explicitreducedsoln}
\end{eqnarray}

Considering first the purely bosonic case, we see that we can take $V_0 = I_N$ and $U_0 = I_M$. This implies that the purely bosonic part of $L_{R}$ is $L_{0R} = L_0$. Note that now the 
unitarity of $U$ implies that 
\begin{eqnarray}
U U^{\dagger} = U^{\dagger} U = I_M \iff U_1 = - U_2^{\dagger}\,, 
\quad  U_3^{\dagger} + U_3 +  U_1 U_1^{\dagger} +  U_1^{\dagger} U_1 = 0\,.
\label{cononUspecial}
\end{eqnarray}
Moreover, by choosing $U_3^{\dagger} = U_3$, 
we reduce the additional freedom in $U$ and get an explicit form of $U_3$ in terms 
of $U_1$ as
\begin{eqnarray}
U_3 = -\frac{1}{2} \big( U_1 U_1^{\dagger} +  U_1^{\dagger} U_1 \big)\,.
\label{U3}
\end{eqnarray}

It is important to observe here that the remaining gauge freedom is now reduced to the arbitrariness of 
choosing $U_1$.

Using the expressions of $V_0$ and $U$, we can now express the system 
(\ref{explicitreducedsoln}) as a set of 3 matrix equations by identifying 
the coefficients of $\theta_-$, $\theta_+$ and $\theta_+\theta_-$, respectively, 
\begin{eqnarray}
&Z L_2 + Z L_0 U_2 = Z L_{2R}\,, \label{reducecompatibility1} \\
&Z L_1 + Z L_0 U_1 + A L_0 = Z L_{1R} + A_R L_0\,, \label{reducecompatibility2} \\
&Z L_3 + Z L_0 U_3 + Z L_1 U_2 + Z L_2 U_1 + A L_2 + A L_0 U_2 = Z L_{3R} + A_R L_{2R}\,.
\label{reducecompatibility3}
\end{eqnarray}

Since $Z$ is in the MacFarlane parametrization (\ref{zMF}), each of these equations can be 
split into two by taking into account the expressions for $A$ and $A_R$
\begin{eqnarray}
A = \left( \begin{array}{c}
\alpha \\
\beta
\end{array} \right) \,, \quad 
A_R = \left( \begin{array}{c}
\alpha_R \\
\beta_R
\end{array} \right)\,,
\qquad 
\begin{array}{c}
\alpha, \,\, \alpha_R \in \field{C}^{M \times M}\,,  \\
\,\,\,\,\,\,\,\,\,\beta, \,\, \beta_R \in \field{C}^{(N - M) \times M}\,.
\end{array}
\label{AandredA}
\end{eqnarray}
The final form of our equations (\ref{reducecompatibility1})-(\ref{reducecompatibility3}) is then 
\begin{eqnarray}
&L_{2R} = L_2 - L_0 U_1^{\dagger}\,,  \label{reducesplit1} \\
&L_{1R} = L_1 + L_0 U_1 + (\alpha - \alpha_R) L_0\,, \label{reducesplit2} \\
&\beta_R = \beta - K (\alpha - \alpha_R)\,, \label{reducesplit3} \\
&L_{3R} = L_3 + L_0 U_3 + L_2 U_1 - L_1 U_1^{\dagger} - \alpha_R L_{2R} - \alpha L_0 U_1^{\dagger}\,.
\label{reducesplit4}
\end{eqnarray}

Since (\ref{reducesplit3}) does not involve $U_1$, the matrix function $\beta$ cannot 
be modified further by using the gauge freedom. The only equation involving $U_1$ and $\alpha$ ,
in a simple way, is (\ref{reducesplit2}). Hence, assuming $L_1 = L_{1R} = 0$ we get 
\begin{eqnarray}
U_1 = -L_0^{-1} (\alpha - \alpha_R) L_0\,. 
\label{matU1} 
\end{eqnarray}

This enables us to put $\alpha_R = 0$ in the simplified form of $A_R$ which greatly simplifies the 
arbitrariness of $A$ and hence of the solution of the susy $G(M,N)$ $\sigma$-model. Now, we have that
\begin{eqnarray}
U_1 = -L_0^{-1}  \alpha L_0\,,  
\label{U1final}
\end{eqnarray}
together with 
\begin{eqnarray}
\beta_R = \beta - K \alpha\,.  
\label{betafinal}
\end{eqnarray}
We see that by now, the gauge freedom has been completely used up and 
the explicit forms of $L_{2R}$ and $L_{3R}$ can be obtained from (\ref{reducesplit1}) and (\ref{reducesplit4}), 
respectively.

Hence, having started from a solution (\ref{MFparasusy}) in which $A$ is completely arbitrary we reach 
a simplified solution
\begin{eqnarray}
W_R = Z + i \theta_+ \eta A_R\,,  
\label{reducedsummary}
\end{eqnarray}
with 
\begin{eqnarray}
A_R = \left( \begin{array}{c}
0 \\
\beta - K \alpha
\end{array} \right) \,, 
\label{redAwithalpha0}
\end{eqnarray}
and this result holds in the gauge (\ref{U1final}). So the simple form of the solution holds in this particular gauge.

\section{Constant curvature solutions and gauge invariance \label{examples}}
In our analysis of different sets of holomorphic solutions of the susy $G(M,N)$ $\sigma$-models we have already found some of them generalising the purely bosonic ones. We consider now two examples
in order to emphasize the importance and usefulness of the gauge invariance in the context of 
susy $G(M,N)$ $\sigma$-model.

\subsection{Constant curvature solutions of $\field{C}P^{N-1}$ model \label{cpn}}
First we look at such solutions for the case of $M=1$ {\it i.e.}; of the  $\field{C}P^{N-1}$ model.

It has been shown that the susy holomorphic constant curvature solution of 
$\field{C}P^{N-1}$ model can be written in the form \cite{MLthesis}
\begin{eqnarray}
\widetilde{W}(x_+, \theta_+) = u(x_+) + i \theta_+ \eta(x_+) A(x_+)\,,
\label{examplecpnsoln}
\end{eqnarray}
where $u(x_+) =  (u_n(x_+))$ is the Veronese sequence 
with 
\begin{eqnarray}
u_n(x_+) = \sqrt{\left(\begin{array}{c}
N-1 \\
n \end{array}\right)}\, x_+^n\,, 
\quad n = 0,1,2, \ldots, N-1\,,
\label{exampleveroneseseq}
\end{eqnarray}
and $A(x_+) = (a_n(x_+))$, with 
\begin{eqnarray}
a_n(x_+) = -a_0(x_+)(n-1)u_n(x_+) + \frac{a_1(x_+)}{\sqrt{N-1}} \partial_{+} u_n(x_+)\,, 
\quad n = 0,1,2, \ldots, N-1
\,.
\label{ancpn}
\end{eqnarray}

We want to simplify this solution by applying to it a gauge transformation $U$, which is 
now a complex function of $(x_{\pm}, \theta_{\pm})$. Indeed, 
\begin{eqnarray}
U = 1 + i \theta_+ \eta {\tilde u}(x_{\pm}) + 
i \theta_- \eta^{\dagger} {\tilde u}^{*}(x_{\pm}) - \theta_+\theta_-\eta^{\dagger}\eta |{\tilde u}(x_{\pm}) |^2\,.
\label{exampledecomU}
\end{eqnarray}

From our preceding discussion, we see that the choice of the gauge $\tilde u$ is 
\begin{eqnarray}
{\tilde u} = \alpha_R - \alpha\,,
\label{examplechoicegauge}
\end{eqnarray}
since $L_0$ is a scalar function. Due to the fact that we want $\alpha_R = 0$, we see that
\begin{eqnarray}
{\tilde u} = - \alpha = -a_0(x_+)\,.
\label{examplefinalcpngauge}
\end{eqnarray}

Our ultimate goal is to find the expression for $A_R$ which appears in $\widetilde{W}_R$. 
Indeed, solving the equation 
\begin{eqnarray}
\beta_R = \beta - K \alpha\,,
\label{excpnbetaalpha}
\end{eqnarray}
in our special case, we get the final form of $A_R$ as 
\begin{eqnarray}
a_{Rn} = \frac{n}{\sqrt{N-1}} \frac{u_n(x_+)}{x_+} \big(a_1(x_+) - a_0(x_+) \sqrt{N-1}\,\,x_+ \big)\,,
\qquad
n = 0, \ldots, N-1\,. 
\label{aRncpn}
\end{eqnarray}

This means that the unique susy holomorphic constant curvature solution of 
$\field{C}P^{N-1}$ model is 
\begin{eqnarray}
\widetilde{W}(x_+, \theta_+) = u(x_+) + i \theta_+ \frac{\tilde{\eta}(x_+)}{\sqrt{N-1}}  \partial_{+} u_n(x_+) \,, 
\qquad
n = 0, \ldots, N-1\,, 
\label{examplepreviousarticle}
\end{eqnarray}
where $\tilde{\eta}(x_+) = \big(a_1(x_+) - a_0(x_+) \sqrt{N-1}\,\,x_+ \big)$ is an arbitrary function of $x_+$,  in agreement with our previous result \cite{DHYZ}. That is the solution 
(\ref{examplepreviousarticle}) given in our previous theorem \cite{DHYZ} is unique up to 
gauge transformations.

\subsection{The simplest holomorphic constant curvature solution of $G(2,4)$ $\sigma$-model \label{g24}} 
Considering the special solution $W_1$ of the form (\ref{MFparasusy}) with 
\begin{eqnarray}
Z_1 = \left( \begin{array}{c}
I_2 \\
K_1
\end{array} \right) \,, \qquad 
K_1 = \left( \begin{array}{cc}
x_+ & 0 \\
0 & 0
\end{array} \right) \,,
\label{exampleZ1}
\end{eqnarray}
it can be shown that $A$ takes the form \cite{MLthesis}
\begin{eqnarray}
A_1 = \left( \begin{array}{c}
\alpha \\
\beta
\end{array} \right) =
\left( \begin{array}{cc}
\alpha_{11} (x_+) & \alpha_{12} (x_+) \\
\alpha_{21} (x_+) & \alpha_{22} (x_+) \\
\beta_{11} (x_+) & c_0 + \big( c_1 + \alpha_{12} (x_+) \big)x_+ \\
b_0 + b_1 x_+ & d_0
\end{array} \right) \,,
\label{exampleformA1}
\end{eqnarray}
in order to have a constant curvature solution. 

This solution $W_1$, can be reduced by using the gauge invariance (indeed using only $U_1$ given in (\ref{U1final})) 
to the following form
\begin{eqnarray}
W_{1R} = Z_1 + i \theta_+ \eta(x_+) A_R(x_+)\,,
\label{exampleW1}
\end{eqnarray}
with
\begin{eqnarray}
A_R = \left( \begin{array}{c}
0 \\
\beta_R - K_1 \alpha
\end{array} \right) \,,  \qquad 
\beta_R = \left( \begin{array}{cc}
\beta_{11} (x_+) - \alpha_{11} (x_+) x_+ & c_0 + c_1 x_+ \\
b_0 + b_1 x_+ & d_0
\end{array} \right) \,.
\label{redAexample}
\end{eqnarray}

\section{Conclusions and final comments \label{conc}}
In this article we have shown that the gauge invariance of the 
susy $G(M,N)$ $\sigma$-model can be used effectively to obtain 
simple forms of the holomorphic constant curvature solutions. Among other things 
this has allowed us to claim that: 
\begin{itemize}
\item All solutions that can be obtained by direct calculations ({\it e.g.}; the 
examples of the constant curvature solutions for $\field{C}P^{N-1}$ $\sigma$-model and for $G(2,4)$ $\sigma$-model) 
could be written in such a simplified form.
\item In order to solve the general problem (say for $G(2,4)$), the computations can be started 
by assuming $A$ to be of the form (\ref{redAwithalpha0}).
\end{itemize}

We believe that exploiting the gauge invariance in such an effective way many properties 
of the solutions of the susy $G(M,N)$ $\sigma$-model can be understood more easily. In particular, 
work is in progress on determining all constant curvature solutions of the susy $G(2,4)$ $\sigma$-model
in which all holomorphic bosonic solutions are given by: 
\begin{equation}
\begin{split}
&Z_1=\left[\begin{array}{cc}
1&0\\
0&1\\
x_+&0\\
0&0\\
\end{array}\right], \  Z_2=\left[\begin{array}{cc}
1&0\\
0&1\\
x_+^2 \cos{2t}&\sqrt{2}x_+\cos{t}\\
\sqrt{2}x_+\sin{t}&0\\
\end{array}\right]\,,  \\[3ex]
&Z_3=\left[\begin{array}{cc}
1&0\\
0&1\\
\sqrt{3}x_+^2&\sqrt{8/3}x_+ \\
0&\sqrt{1/3}x_+\\
\end{array}\right], \  Z_4=\left[\begin{array}{cc}
1&0\\
0&1\\
2x_+^3&\sqrt{3}x_+^2\\
\sqrt{3}x^2&2x_+\\
\end{array}\right]\,,
\end{split}
\label{Z1234}
\end{equation}
where $t$ is a real parameter. Up to a $U(4)$ gauge transformation, these are the 
only bosonic solutions with constant curvature \cite{4solutionsG(24)}. 

Finally, the next step 
would be to analyse the constant curvature solutions of the susy $G(2,5)$ and in general  
the susy $G(2,N)$ $\sigma$-models where the purely bosonic solutions are given in 
\cite{solutionsG(25)} and \cite{solutionsG(2n)}, respectively.

\section*{Acknowledgments}
The work of \.{I}Y was partly supported by Hacettepe University Scientific Research 
Coordination Unit. Project Number: FBI-2017-14035. He also thanks the Centre de 
Recherches Math\'{e}matiques, Universit\'{e} de Montr\'{e}al for the kind hospitality 
while on sabbatical leave during which this work was done. VH acknowledges the support of research grants from NSERC of Canada.


\end{document}